# Superconducting Metamaterials

Michael Ricci, Nathan Orloff and Steven M. Anlage

*Center for Superconductivity, Physics Department, University of Maryland, College Park, MD 20742*

Evanescent wave amplification has been predicted under the ideal condition that the index of refraction, $n = -1 + i\, 0$ precisely, but is difficult to observe in practice because current metamaterials suffer from high losses. We present experimental results on a metamaterial that employs superconducting Nb metals and low-loss dielectric materials. Results include transmission data on a wire, split-ring resonator, and a combination medium at temperatures between 4.2 K and 297 K. Evidence of negative effective permittivity, permeability, and a negative effective index passband are seen in the superconducting state between 50 MHz and 18 GHz. We find a dielectric loss of $\varepsilon_{eff,2} = 2.6 \times 10^{-3}$ in a superconducting wire array at 10.75 GHz.

There has been renewed interest in the properties of materials with a negative index of refraction (NIR), originally proposed in 1967 by Veselago.[1] Compelling experimental evidence for this behavior has been seen through negative index passbands,[2-5] negative refraction,[3,6] and apparent super-resolution imaging.[7-10] These demonstrations have been made at room temperature in artificial loop-wire,[2] dual circuit,[11] and photonic crystal[12] media.

An important prediction for NIR materials is evanescent wave amplification under the ideal condition of $n = -1 + i\, 0$ precisely.[14] This property will permit, in principle, image reconstruction with arbitrary precision and detail as demonstrated, for example, in calculations of flat-lens transfer functions.[13-15] However, these theoretical works also show that ideal evanescent wave amplification suffers from three important constraints; the real part of $n$ must be -1, exactly, the metamaterial must be thin (compared to the wavelength) to minimize retardation effects, and the imaginary part of $n$ must be much less than 1, so that there is very little damping. The first two constraints can be satisfied, in principle, with appropriate engineering of existing metamaterials. However, significant losses have a debilitating effect on most designs.

The loss constraint is illustrated through calculations of the transfer function through a flat lens. The p-polarized light transfer function through a thin $\varepsilon_{eff} = -1 + i\, \varepsilon_{eff,2}$ lens of thickness $d$ is approximately,[14]
$$T_p \cong e^{-k_x d}(i\varepsilon_{eff,2}/2 + e^{-2k_x d})^{-1},$$
where $k_x$ is the lateral wavenumber. Exponential amplification requires $\varepsilon_{eff,2} \ll e^{-2k_x d}$, which constrains both the upper spatial frequency that can be amplified and the slab thickness. For example, to amplify a lateral wavenumber with $k_x/k_0 = 2$ through a slab of thickness $d/\lambda = 0.1$ requires $\varepsilon_{eff,2} \ll 8 \times 10^{-3}$. Such low losses are only available under special circumstances.

In this paper we consider the use of superconducting (SC) metals and low-loss dielectrics. These metamaterials reveal important properties, in addition to satisfying the third constraint mentioned above. We consider Nb wires (for negative effective permittivity, $\varepsilon_{eff}$), and Nb split-ring resonators (SRRs, for negative effective permeability, $\mu_{eff}$) in a quasi-one-dimensional waveguide geometry.[4] The results are compared to predictions for the plasma edge of an infinite wire array,[16,17] and the magnetic properties of SRRs.[18]

A Ag-plated X-band (WR90) waveguide (interior dimensions 22.86 mm by 10.16 mm) is employed for the cryogenic transmission measurements. The top and bottom walls of the 10-cm-long waveguide are perforated with 0.51-mm-diameter holes (5 holes wide, 21 rows long), and cylindrical 0.25-mm-diameter SC Nb wire is threaded through the holes to make a vertical wire array (Fig. 1, inset). Various periodic wire arrays can be created, including an $a = 5.08$ mm and a 45°-rotated $a = 7.19$ mm square array. The Nb wire has a SC transition temperature ($T_c$) of 9.25 K, as measured by AC susceptibility.

The SRRs are made from 200-nm-thick Nb thin films RF-magnetron sputtered onto single crystal quartz substrates. The square SRRs ($T_c$ of 8.65 K) are photolithographically patterned and have a separation and gap width of 300 $\mu$m, a line width of 154 $\mu$m, an inner ring width of 1448 $\mu$m, and an outer ring width of 2360 $\mu$m, similar in design to Refs. [3,19]. The 350-$\mu$m-thick quartz wafers are cut into chips 9.0 mm tall by 44.8 mm long, and placed into the waveguide to couple with the transverse magnetic fields of the microwave signals (Fig. 2, inset). The SRR gaps are oriented orthogonal to the electric field. In all cases, the SRR elements, wire diameters, and lattice spacing are all much smaller than the wavelength of the signals, creating an effective medium.

The waveguide assembly is placed in a magnetically shielded cryogenic Dewar and connected to an Agilent 8722D vector network analyzer through waveguide/coaxial couplers and nearly 1.37 meters of coaxial cable. The 2 by 2 complex S-matrix (including the transmission coefficient $S_{21}(f)$) of the waveguide is measured between 50 MHz and 18 GHz. We compare these measurements to a waveguide wave matrix model,[20] using effective medium models for the permittivity and permeability, and fit the data with a nonlinear least-squares fit.

We first examined the plasma edge properties of a SC wire array with the dominant $TE_{10}$ mode (electric field parallel to wires) for two different lattice parameters. Transmission data for the $a = 7.19$ mm lattice are shown in Fig. 1, at two temperatures, 7 K (below the wire $T_c$),

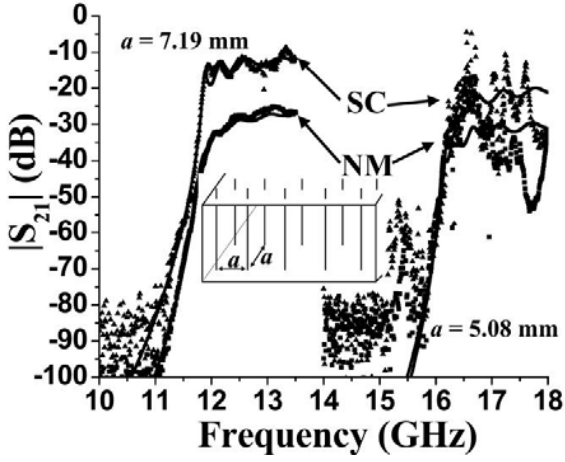

Fig. 1 Measurements of transmission magnitude |S$_{21}$| versus frequency through Nb wire arrays in an X-band waveguide. Spectra on the left are for a lattice parameter $a$ = 7.19 mm, and those on the right are for $a$ = 5.08 mm. The upper traces (triangles) are taken in the SC state at 7 K, and the lower traces (squares) are taken in the NM state at room temperature. Theoretical fits to the data are solid lines. The inset shows a diagram of the wire array looking in the propagation direction.

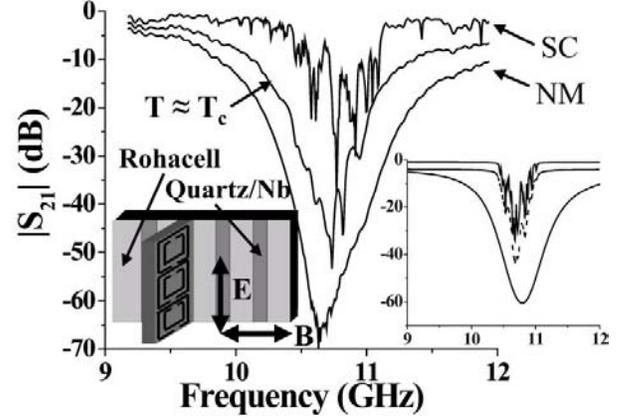

Fig. 2 Measurements of transmission magnitude |S$_{21}$| versus frequency through a Nb/Quartz SRR array in an X-band waveguide. The upper trace is taken in the SC state at 5.0 K, the middle trace is taken near T$_c$ of the SRRs, and the lower trace is taken in the NM state at 9.0 K. The left inset shows a diagram of the SRR array looking in the propagation direction. The right inset shows theoretical models in the SC state (upper curve), close to T$_c$ (dashed), and NM state (lower curve).

and 297 K.

It is readily seen that the insertion loss is reduced in the SC state, and the plasma edge is sharper. (All data presented here have been corrected for the insertion loss of the coaxial cables.) Fits using an effective dielectric response[16] of $\varepsilon_{eff}(\omega) = 1 - \omega_p^2/\omega(\omega + i\gamma)$ are also shown in Fig. 1. The loss parameters ($\gamma/2\pi$) for the SC and normal metal (NM) states were found to be 24 MHz and 146 MHz, respectively, illustrating the reduction of loss in the SC state. We find $\varepsilon_{eff,2}$ = 2.6 x 10$^{-3}$ (SC), and $\varepsilon_{eff,2}$ = 1.6 x 10$^{-2}$ (NM), below the plasma edge at 10.75 GHz.

Pendry predicts the plasma edge of an infinite three-dimensional cubic wire network to be given by[16] $\omega_p^2 = 2\pi c^2(a^2(\ln(a/r\sqrt{\pi}) + \pi r^2/2a^2 - 1/2))^{-1}$, where $c$ is the speed of light, $a$ the lattice parameter, and $r$ the wire radius. Efros and Pokrovsky have a different prediction discussed in Ref. [17]. The fit plasma frequencies for this lattice in the SC (NM) state are 11.785 GHz (11.728 GHz), which is somewhat larger than the predicted values of 9.64 GHz[16] and 9.95 GHz.[17]

Transmission data through the $a$ = 5.08 mm lattice are also shown in Fig. 1 at 7 K and 291 K. The plasma frequency increases as the wire density increases, as expected. This lattice has a fit plasma frequency of 16.135 GHz (16.028 GHz) in the SC (NM) state, and predicted values of 14.52 GHz[16] and 16.9 GHz.[17] The results for this lattice are less clear because at least two higher order modes are excited in the waveguide above 15 GHz. The model does not account for these modes, and so $\gamma/2\pi$ is found to be the same for both (~ 10 MHz), although the insertion loss decreases in the SC state. In all cases, no sudden changes in the transmission properties are seen at the wire T$_c$, and there is no evidence of enhanced transmission below a lower resonant frequency down to 50 MHz.

The results of the Nb/quartz SRR arrays in the waveguide are shown in Fig. 2. Four strips, two chips long (eighteen columns of three SRRs) were placed in the waveguide, spaced by Rohacell 31 HF ($\varepsilon_r$ = 1.046 at 10 GHz), as shown in the left inset. The SRRs are expected to resonate in the vicinity of 10.75 GHz, where one expects an $\varepsilon_{eff}$ > 0, $\mu_{eff}$ < 0 transmission notch. Note that in the NM state (9.0 K), the SRRs act collectively, giving a smooth notch centered at the expected frequency of 10.76 GHz. However, at 5.0 K (below T$_c$), the SRRs seem to act individually, and give a distribution of relatively high quality factor resonant frequencies, presumably determined by a distribution of local environments and interactions.[21]

These results are modeled well assuming that $\mu_{eff}$[18] takes on an average value given by the superposition, $\bar{\mu}_{eff} = N^{-1}\sum_1^N \mu_{eff}(f, f_{0i}, \Gamma, F)$, where $f_{0i}$ is the $i^{th}$ resonant frequency, $F$ and $\Gamma$ are the (assumed) global magnetic filling fraction and loss parameter. Based on the data, we take a Gaussian distribution (variable center frequency, and fixed width of 0.13 GHz) of $N$ = 108 resonant frequencies in the model. These model results are shown in the right inset of Fig. 2. Depending on the loss parameter, $\Gamma$, a smooth or jagged curve is obtained. The corresponding values of $\Gamma/2\pi$ for the SC state and the NM state are 13 MHz and 637 MHz, respectively, and 48 MHz for the temperatures near T$_c$. Although this is an indirect way to determine $\mu_{eff, 2}$, estimates extracted from these calculations vary from $0.12 \leq \mu_{eff, 2} \leq 1.0$ (mean value 0.41) in the NM state, and $3.6 \times 10^{-4} \leq \mu_{eff, 2} \leq 0.4$ (mean value 0.067) in the SC state.

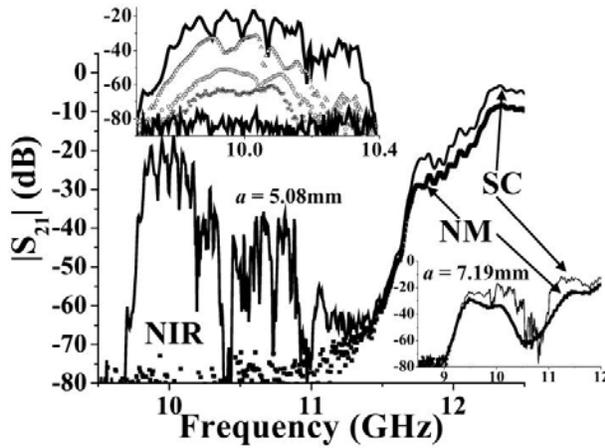

Fig. 3 Measurements of transmission magnitude $|S_{21}|$ versus frequency through a composite 90-mm-long Nb wire and Nb/Quartz SRR array in an X-band waveguide with $a$ = 5.08 mm. The upper trace (solid line) is taken in the fully SC state at 5.5 K, and the lower trace (squares) is taken in the fully NM state at 11 K. The left inset shows a close-up of the NIR region at 7.0, 8.08, 8.34, 8.49, and 8.77 K, from top to bottom (SRR $T_c$ = 8.65 K). The right inset shows data from the composite with $a$ = 7.19 mm in the SC (solid line) and NM (squares) states.

Building upon these results, a SC metamaterial was created by combining a Nb wire array, with Nb SRR strips supported by Rohacell in the middle of, and centers aligned with, the columns of wires. Fig. 3 contains the results of transmission measurements. The plasma edge of the $a$ = 5.08 mm wire array has been significantly reduced by the presence of the Nb/Quartz SRR wafers, consistent with the results of Ref. [22]. In this case, the NIR peak is apparent in the SC state between 9.5 and 11 GHz, consistent with the $\mu_{eff} < 0$ notch frequency seen in the SRRs-only measurements (Fig. 2). Several observations support this interpretation. The left inset shows a close-up of the NIR peak as the temperature increases. As the temperature approaches the SRR film $T_c$, the peak becomes smooth, consistent with Fig. 2, and decreases in magnitude as the losses become significant. Most previous work has utilized thick (~30 $\mu$m) Cu plating, where $\Gamma/2\pi \sim$ 1 MHz, which is much less than our estimate ($\Gamma/2\pi \sim$ 1 GHz[18]) for the Nb thin film in the NM state just above $T_c$. It is clear that the high loss of the NM thin film SRRs prevents the observation of a NIR peak. In addition, the edge near 11.5 GHz does not change at either the film or wire $T_c$, consistent with it being the plasma edge of the composite system.

To further support our picture, the right inset shows data for the $a$ = 7.19 mm wire array and SRRs. It appears that the plasma edge is reduced to just above 9 GHz, and now the dip centered near 10.5 GHz is due to the SRR resonance. That is, we have a $\mu_{eff} < 0$ region *above* the plasma edge of the wires, and thus we *do not* have a NIR peak. In fact, below $T_c$ the dip is jagged, and above $T_c$ the dip is smooth, very similar to the $\varepsilon_{eff} > 0$, $\mu_{eff} < 0$ data in Fig. 2. From these results it is clear that the elements of the composite metamaterial are strongly interacting, and its properties are not simply a superposition of the wires and SRRs separately.[22]

We have created and examined the properties of a low-loss SC metamaterial. It displays sharp plasma edges with SC wires, and exhibits features in the SRR properties not seen in normal metals. Preliminary estimates show that $\varepsilon_{eff, 2}$ and $\mu_{eff, 2}$ have both been reduced by a factor of about 6, upon entering the SC state. We also demonstrate NIR behavior through superposition of $\varepsilon_{eff} < 0$, $\mu_{eff} < 0$ contributions in a SC composite metamaterial with a wire lattice parameter of $a$ = 5.08 mm. The present design shows substantial variation of the SRR resonant frequencies, even in a uniform array.

This work is supported by the National Science Foundation through grant NSF/ECS-0322844. We gratefully acknowledge assistance from R. Frizzell, J. Hamilton, K. Mercure, A. Prasad, and C. P. Vlahacos.